 \documentclass[final,5p,times,twocolumn]{elsarticle}

\usepackage{amssymb}
\usepackage{lipsum}
\usepackage{amsmath}
\usepackage{cuted}
\usepackage{doi}
\usepackage[utf8]{inputenc}
\usepackage{textgreek}

\usepackage{listings}

\journal{Physics Letters B}

\begin{document}

\begin{frontmatter}



\title{A practical methodology for $\Lambda$ global polarization extraction in fixed-target experiments}


\author[first]{Tan Lu}
\affiliation[first]{organization={State Key Laboratory of Heavy Ion Science and Technology, Institute of Modern Physics, Chinese Academy of Sciences},
            city={Lanzhou},
            postcode={730000}, 
            country={China}}

\author[first]{Chengdong Han}

\author[second]{Chenlu Hu}
\affiliation[second]{organization={University of Chinese Academy of Sciences},
            city={Beijing},
            postcode={100049}, 
            country={China}}

\author[first]{Xionghong He}

\author[first]{Diyu Shen \corref{cor1}}
\ead{dyshen@impcas.ac.cn}
\cortext[cor1]{Corresponding author}

\author[first]{Subhash Singha \corref{cor1}}
\ead{subhash@impcas.ac.cn}

\author[third]{Shusu Shi}
\affiliation[third]{organization={Central China Normal University},
            city={Wuhan},
            postcode={430079}, 
            country={China}}

\author[third]{Xing Wu}

\author[second]{Guannan Xie}


\author[first]{Yapeng Zhang}

\begin{abstract}
Non-central heavy-ion collisions generate large orbital angular momentum in the created medium, which leads to polarization of final-state particles via spin–orbit coupling, known as global spin polarization. The observation of significant global polarization of $\Lambda$ hyperon in heavy-ion collisions indicates that the quark–gluon plasma is the most vortical fluid known in nature. Exploring $\Lambda$ global polarization at lower energies is important for understanding spin dynamics across different regions of the quantum chromodynamics (QCD) phase diagram. Low-energy nuclear experiments are typically conducted with asymmetric detector acceptance, as in fixed-target collisions at RHIC-STAR, and at facilities such as FAIR, NICA, HIAF and HIRFL-CSR. The asymmetric rapidity coverage in these experiments enhances the coupling between directed flow and detector inefficiencies, creating significant bias in $\Lambda$ global polarization measurements. In this paper, we propose a methodology to eliminate such bias arising from asymmetric detector acceptance. The method is validated using realistic detector simulations based on the STAR fixed-target configuration.
\end{abstract}


\end{frontmatter}




\section{Introduction}
In non-central heavy-ion collisions, the produced fireball can carry a fraction of the large orbital angular momentum transferred from the incident nuclei~\cite{Becattini:2007sr}; particles with non-zero spin become polarized along the direction of the orbital angular momentum via spin–orbit coupling, known as global spin polarization~\cite{Liang:2004ph,Liang:2004xn,Voloshin:2004ha}. The weak decay of the $\Lambda$ hyperon violates parity, enabling its polarization to be determined from the angular distribution of its decay products. Measurements of hyperon global spin polarization provide a novel probe of the properties and spin dynamics of the matter produced in heavy-ion collisions.

A major breakthrough was achieved in the first phase of the STAR Beam Energy Scan (BES-I) program, where non-zero global spin polarization of $\Lambda$ and $\bar{\Lambda}$ was observed in Au+Au collisions~\cite{STAR:2017ckg}, indicating that the quark–gluon plasma (QGP) is the most vortical fluid known in nature. Subsequently, more comprehensive and differential measurements of global and local polarization of $\Lambda$ and $\bar{\Lambda}$ hyperons have been performed over a wide range of collision energies in heavy-ion experiments at RHIC~\cite{STAR:2018gyt,STAR:2020xbm,STAR:2021beb,STAR:2023nvo,STAR:2020xbm,STAR:2023qyt,STAR:2025dgs,STAR:2019erd,STAR:2023eck}, the LHC~\cite{ALICE:2019onw,ALICE:2021pzu}, and HADES~\cite{HADES:2022enx}. Similar global polarization signals are also observed for multi-strange hyperons such as $\Xi$ and $\Omega$~\cite{STAR:2020xbm,ALICE:2021pzu}, supporting the global nature of the vortical structure across different hyperon species. Hydrodynamic and transport model calculations reproduce the main features of global polarization over a broad range of collision energies~\cite{Becattini:2024uha,Chen:2024aom,Karpenko:2016jyx,Vitiuk:2019rfv,Voronyuk:2023vyu,Ivanov:2019ern,Ivanov:2020udj,Ivanov:2025izv,Li:2017slc,Fu:2020oxj,Guo:2021udq,Becattini:2020ngo,Huang:2020dtn,Becattini:2021lfq}.

Most measurements of hyperon global polarization in collider experiments are performed under approximately symmetric rapidity acceptance, where the directed flow averaged over rapidity is close to zero. In this framework, the polarization is determined from correlations between the angular distribution of hyperon decay products and the reaction plane orientation, which reflects the direction of the system orbital angular momentum. The experimental formalism for extracting global polarization, including its connection to anisotropic flow techniques and event-plane reconstruction~\cite{Poskanzer:1998yz,Voloshin:2008dg,Niida:2024ntm}. It was developed by the STAR Collaboration~\cite{Selyuzhenkov:2006tj,STAR:2007ccu}. In collider detectors with nearly symmetric rapidity coverage, contributions from directed flow largely cancel, allowing the global polarization to be extracted using this approach.

Exploring global spin polarization in the high-baryon-density region of the QCD phase diagram is crucial for understanding spin–orbit coupling in strong interactions. In relativistic hydrodynamic approaches, the global polarization of $\Lambda$ is related to the thermal vorticity $\omega/T$ under local thermal equilibrium~\cite{Becattini:2012tc,Becattini:2013fla}, while such proportionality may be modified at sufficiently low collision energies where the system evolution deviates from equilibrium conditions. On the other hand, the total orbital angular momentum of the system depends on the collision geometry and beam energy, approximately scaling as $J_y \sim Ab\sqrt{s_{NN}-2m_N^2}/2$, where $A$ and $b$ denote the nuclear mass number and impact parameter, respectively~\cite{Akridge:2025jgy}. As the collision energy decreases toward the production threshold, the available angular momentum carried by participant matter is reduced, while baryon stopping becomes increasingly important. The competition between these effects may lead to a non-monotonic dependence of global polarization on collision energy. Model calculations based on hydrodynamic, transport, and initial-state approaches indicate that the $\Lambda$ global polarization, or the angular momentum deposited in the midrapidity region, may exhibit a maximum in the few-GeV energy range, although the exact location of the peak varies among different theoretical descriptions~\cite{Deng:2020ygd,Deng:2021miw,Ivanov:2020udj,Guo:2021udq,Ivanov:2022ble}. Experimental programs targeting this energy region include the STAR fixed-target program at RHIC ($\sqrt{s_{NN}}=$ 3 – 13.7 GeV), as well as experiments at FAIR (CBM, HADES)~\cite{Friman:2011zz,Almaalol:2022xwv}, NICA~\cite{Voronyuk:2023vyu}, HIRFL-CSR (CEE) and HIAF~\cite{Lu:2016htm,Hu:2023niz,Xia:2002xpu,Mao:2020rlb,Zhou:2022pxl,Yang:2025sni}, extending measurements toward the $\Lambda$ sub-threshold region. In such fixed-target configurations, the rapidity acceptance becomes intrinsically asymmetric, leading to non-zero averaged directed flow within the detector coverage. Therefore, an analysis method that properly accounts for asymmetric detector acceptance is required.

In this paper, we propose a data-driven method for measuring global polarization in fixed-target experiments that accounts for finite directed flow. The method is demonstrated using $\Lambda$ hyperons and can be extended to multi-strange hyperons such as $\Xi$ and $\Omega$ with appropriate treatment of their decay kinematics. This paper is organized as follows. In Sec.~II, we derive the new observables analytically. In Sec.~III, we employ the STAR GEANT framework~\cite{Brun:1994aa,Fine:2000,STAR:2001rbj} to simulate realistic detector response and validate the method. In Sec.~IV, we summarize the results and discuss the outlook for future experimental studies.

\section{Methodology}
Parity violated weak decay of $\Lambda$ hyperons has the following differential cross section owing to $S$ wave and $P$ wave superposition,
\begin{equation}
    \frac{dN_p}{d\Omega^*} \propto  1+\alpha_\Lambda P_\Lambda\cos(\theta^*),  
\label{eq:decay_distribution}
\end{equation}
where $\alpha_\Lambda$ is a decay constant which is measured to be $0.732 \pm 0.014$, reflecting the magnitude of parity violation~\cite{ParticleDataGroup:2022pth}. The $\theta^*$ is the angle between polarization direction and momentum direction of daughter proton in $\Lambda$ rest frame. 

In heavy-ion experiments, the global polarization is measured along the direction perpendicular to the reaction plane which is defined by the impact parameter and beam direction. The reaction plane determination in experiments has finite resolution in transverse direction, therefore experimental observables are usually constructed based on the azimuthal projection of Eq.~\ref{eq:decay_distribution}:
\begin{equation}
    \frac{dN_p}{d\Omega^*} \propto  1+\alpha_\Lambda P_\Lambda\sin(\theta^*_p)\sin(\Psi_1-\phi^*_p) ,
\label{eq:decay_distribution_transverse}
\end{equation}
\begin{figure}[!htbp]
    \centering
    \includegraphics[width=0.6\linewidth]{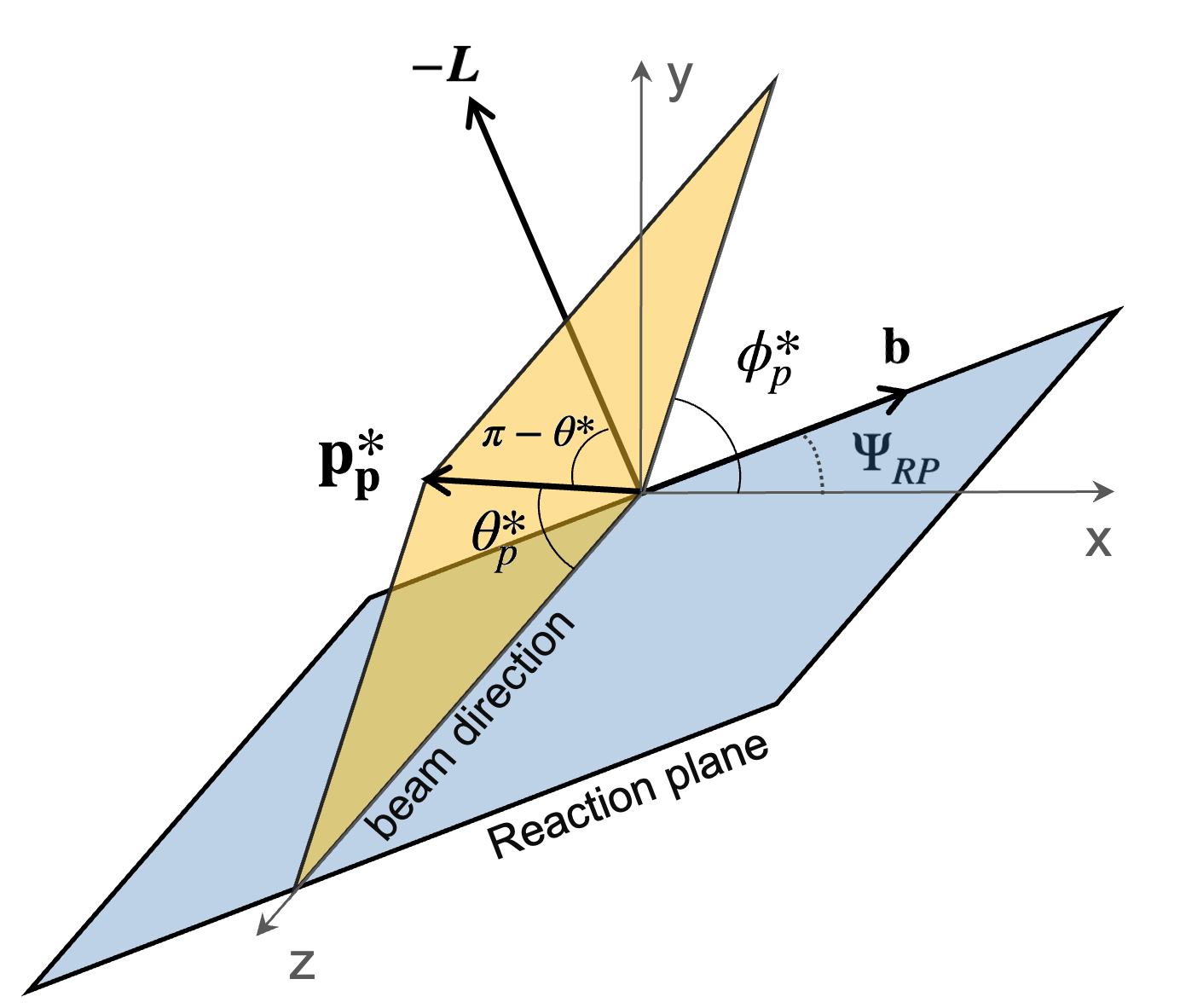}
    \caption{Illustration for angles in global polarization measurement.}
    \label{fig:placeholder}
\end{figure}

where $\Psi_1$ is the first-order event plane, $\theta_p^*$ and $\phi_p^*$ are the polar angle and azimuthal angle of daughter proton in $\Lambda$ rest frame, respectively. From Eq.~\ref{eq:decay_distribution_transverse}, the global polarization can be measured by taking average of $\sin(\Psi_1 - \phi_p*)$ over full space,
\begin{equation}
\begin{aligned}
\langle \sin(\Psi_{1}-\phi_p^*)\rangle &= \frac{\int d\Omega^* \sin(\Psi_{1}-\phi_p^*)\frac{dN_p}{d\Omega^*} }{\int d\Omega^* \frac{dN_p}{d\Omega^*} },  \\
&= \frac{\pi \alpha_\Lambda P_\Lambda}{8}.
\label{eq:fullAccep}
\end{aligned}
\end{equation}
Eq.~\ref{eq:fullAccep} can be applied to an ideal experiments with $4\pi$ acceptance and perfect detection efficiency. Moreover, Eq.~\ref{eq:fullAccep} implicitly assumed isotropic $\Lambda$ azimuthal distribution. However, at low energy heavy-ion collisions, the directed flow is significant and leads to the following distribution,
\begin{equation}
    \frac{dN_\Lambda}{d\phi_\Lambda } \propto  1+2v_1\cos(\phi_\Lambda -\Psi_{1}),
    \label{eq:flow}
\end{equation}
where $\phi_\Lambda$ is the azimuthal angle of $\Lambda$ in laboratory frame. Within finite detector acceptance, the event average of the polarization observable $\sin(\Psi_{1}-\phi_p^*)$ can be analytically expressed as 

\begin{strip}
\begin{equation}
    \begin{aligned}
            \langle \sin(\Psi_{1}-\phi_p^*)\rangle &= \frac{\sum_{i=0}^{N_\Lambda} \sin(\Psi_{1}-\phi_p^*)}{N_\Lambda} = \frac{\int d{\bf p_\Lambda} \int d{\bf p_p^*} \int \frac{d\Psi_{1}}{2\pi} A({\bf p_\Lambda}, {\bf p_p^*)} \frac{dN_p}{d\Omega^*}  \frac{dN_\Lambda}{d\phi_\Lambda}\sin(\Psi_{1}-\phi_p^*)}{\int d{\bf p_\Lambda} \int d{\bf p_p^*}  \int \frac{d\Psi_{1}}{2\pi}  A({\bf p_\Lambda}, {\bf p_p^*)} \frac{dN_p}{d\Omega^*}  \frac{dN_\Lambda}{d\phi_\Lambda}},  \\ 
            &= \frac{\int d{\bf p_\Lambda} \int d{\bf p_p^*}  A({\bf p_\Lambda}, {\bf p_p^*)}\Big [ v_1 \sin(\phi_\Lambda -\phi_p^*) + \frac{\alpha_\Lambda P_\Lambda}{2}\sin(\theta_p^*) \Big ]}{\int d{\bf p_\Lambda} \int d{\bf p_p^*} \int  A({\bf p_\Lambda}, {\bf p_p^*)} \Big [ 1 + v_1\alpha_\Lambda P_\Lambda \sin(\theta_p^*) \sin(\phi_\Lambda -\phi_p^*)  \Big ]}.
            \label{eq:DefinitionofAverage}
    \end{aligned}
\end{equation}
\end{strip}

\leavevmode
The detector acceptance is described by $A({\bf p_\Lambda}, {\bf p_p^*)}$ that depends only on the $\Lambda$ momentum ${\bf p_\Lambda}$, and momentum of proton in $\Lambda$ rest frame, $\bf p_p^*$. Using Eq.~\ref{eq:decay_distribution_transverse} and Eq.~\ref{eq:flow}, one can integrate event plane $\Psi_1$ over $2\pi$. Note that the denominator in Eq.~\ref{eq:DefinitionofAverage} represents overall $\Lambda$ reconstruction efficiency, it depends on the product of $\Lambda$ directed flow and polarization. In the experiments of beam-beam collisions, the detector has symmetric forward and backward acceptance, therefore the overall $v_1$ is zero due to cancellations, and Eq.~\ref{eq:DefinitionofAverage} is reduced to the method which has been widely used in such experiments~\cite{STAR:2018gyt,STAR:2020xbm,STAR:2023nvo,STAR:2020xbm,STAR:2023qyt,STAR:2025dgs,STAR:2019erd,STAR:2023eck,ALICE:2019onw,ALICE:2021pzu}. 

Fixed-target experiments usually have asymmetric rapidity acceptance which leads to non-zero $v_1$. In previous publications~\cite{STAR:2021beb,HADES:2022enx}, the $ \langle \sin(\Psi_{1}-\phi_p^*)\rangle$ was measured at given $\Psi_1 - \phi_p^* = x$, and was fitted with
\begin{equation}
	f = c_1\sin(x) + \frac{\pi \alpha_\Lambda}{8}A_0P_\Lambda,
\end{equation}
where $A_0 = 4/\pi \langle \sin(\theta_p^*) \rangle $. There are two implicit assumptions:

a). It is assumed that the directed flow $v_1$ is independent of $\bf p_\Lambda$, so that it can be parameterized as a constant $c_1$. 

b). The contribution from $v_1P_\Lambda$ product term in $dN_p/d\Omega^* \times dN_\Lambda/d\phi_\Lambda$ was was not considered in the previous treatment. However, in experiments with large $P_{\Lambda}$ and $v_1$, or when high precision is required, this contribution should be taken into account.

We therefore propose a method to eliminate the above biases in polarization measurement. We consider the following quantity
\begin{strip}
\begin{equation}
    \begin{aligned}
            \langle \sin(\phi_\Lambda - \phi_p^*)\cos(\phi_\Lambda - \Psi_1) \rangle 
            &= \frac{\int d{\bf p_\Lambda} \int d{\bf p_p^*}  A({\bf p_\Lambda}, {\bf p_p^*)}\Big[ v_1 \sin(\phi_\Lambda-\phi_p^*) + \frac{\alpha_\Lambda P_\Lambda }{2} \sin\theta_p^* \sin^2(\phi_{\Lambda}-\phi_p^*) \Big]}{\int d{\bf p_\Lambda} \int d{\bf p_p^*} \int  A({\bf p_\Lambda}, {\bf p_p^*)} \Big [ 1 + v_1\alpha_\Lambda P_\Lambda \sin(\theta_p^*) \sin(\phi_\Lambda -\phi_p^*)  \Big ]}.
            \label{eq:SinCos}
    \end{aligned}
\end{equation}
\end{strip}

Comparing Eq.~\ref{eq:DefinitionofAverage} and Eq.~\ref{eq:SinCos}, it is found that the $v_1$ contribution is the same. By taking the difference between Eq.~\ref{eq:DefinitionofAverage} and Eq.~\ref{eq:SinCos}, we have  
\begin{equation}
	\begin{aligned}
	&\langle\sin(\Psi_1-\phi_{p}^*) - \sin(\phi_\Lambda-\phi_p^*) \cos(\phi_\Lambda-\Psi_1)\rangle \\
	&= \frac{1}{D} \int d{\bf p_\Lambda} \int d{\bf p_p^*}  A({\bf p_\Lambda}, {\bf p_p^*)} \frac{\alpha_\Lambda P_\Lambda}{4} \sin\theta_p^* \\ 
    &\quad \times \left[ 1+\cos2(\phi_{\Lambda}-\phi_p^*) \right],
	\end{aligned}
	\label{eq:numerator}
\end{equation}
where $D$ is the denominator of Eq.~\ref{eq:DefinitionofAverage} and Eq.~\ref{eq:SinCos}. Moreover, the average of $\sin\theta_p^*$ can be written as
\begin{equation}
	\begin{aligned}
	    \langle \sin\theta_p^* \rangle 
		& =\frac{1}{D}\int d{\bf p_\Lambda} \int d{\bf p_p^*}  A({\bf p_\Lambda}, {\bf p_p^*})  \sin\theta_p^* \\
		&\quad \times \Big[1 + {v_1 \alpha_\Lambda P_\Lambda } \sin\theta_p^* \sin(\phi_{\Lambda} - \phi_p^*)\Big].
	\end{aligned}
	\label{eq:denominator}
\end{equation}
Taking ratio of Eq.~\ref{eq:numerator} and Eq.~\ref{eq:denominator}, for a given $\phi_\Lambda - \phi_p^*$, we have 

\begin{equation}
	\begin{aligned}
		&\frac{\langle\sin(\Psi_1-\phi_{\mathrm{p}}^*) - \sin(\phi_\Lambda-\phi_p^*)\cos(\phi_\Lambda-\Psi_1)\rangle}{\langle \sin\theta_p^* \rangle  } \\
		&\quad = \frac{c_0 \left[ 1+\cos2(\phi_\Lambda-\phi_p^*)\right]}{1+c_0c_1\sin(\phi_\Lambda-\phi_{p}^*) }\\
	\end{aligned}
	\label{eq:ratio-method}
\end{equation}
where 
\begin{equation}
    \begin{aligned}
        c_0 &= \frac{\alpha_\Lambda P_\Lambda}{4}, \\
        c_1 &=  \frac{\int d{\bf p_\Lambda} \int d{\bf p_p^*}  A({\bf p_\Lambda}, {\bf p_p^*})  4v_1 \sin^2\theta_p^*} {\int d{\bf p_\Lambda} \int d{\bf p_p^*}  A({\bf p_\Lambda}, {\bf p_p^*})  \sin \theta_p^*}.
        \label{eq:parameters}
    \end{aligned}
\end{equation}
The left-hand side of Eq.~\ref{eq:ratio-method} are the new observable which has to be measured at a fixed $\phi_\Lambda - \phi_p^*$, the right-hand side is the fitting function with parameter $c_0$ and $c_1$. The polarization can be obtained from $c_0$ according to Eq.~\ref{eq:parameters}. In the above derivations, we implicitly assume the polarization $P_\Lambda$ is a unknown constant. If the $P_\Lambda$ has a strong dependence on $\Lambda$ momentum, this method needs to be applied in a narrow kinematic window in which the polarization can be treated as a constant.

\section{Embedding data validation}
In this section, we employ STAR GEANT embedding simulation~\cite{STAR:2001rbj,STAR:1997sav,STAR:1999sib,Anderson:2003ur} to verify the method in Au+Au 3 GeV fixed-target collision geometry. 
Monte Carlo (MC) $\Lambda$ signals are embedded into real experimental events, with an embedded fraction of approximately 5\% of the primary tracks, and the combined data and embedded tracks are processed through the standard reconstruction chain. 
A total of 20 million events are used for this study. 
The MC $\Lambda$ are generated with uniform transverse momentum distributions from 0 to 3 GeV/$c$, and uniform pseudorapidity distributions from $-2.5$ to 0.3. $\Lambda$ hyperons are reconstructed through KFParticle package~\cite{Gorbunov2013,Zyzak2016,Kisel:2018nvd,Ju:2023xvg} as in experiments~\cite{STAR:2021beb}. Based on the detector acceptance, we select $\Lambda$ with the following kinematic cuts~\cite{STAR:2021beb}: $p_T^\Lambda\in[0.7,\ 2.0]$ GeV/c, $p_T^p>0.4$ GeV/c, and $p_T^{\pi^-}>0.15$ GeV/c. The rapidity for $\Lambda$ in center-of-mass frame is $y-y_{\rm cm}\in[-1.0,\ 0.2]$. 

\begin{figure}[!htbp]
	\centering
	\includegraphics[width=0.48\textwidth]{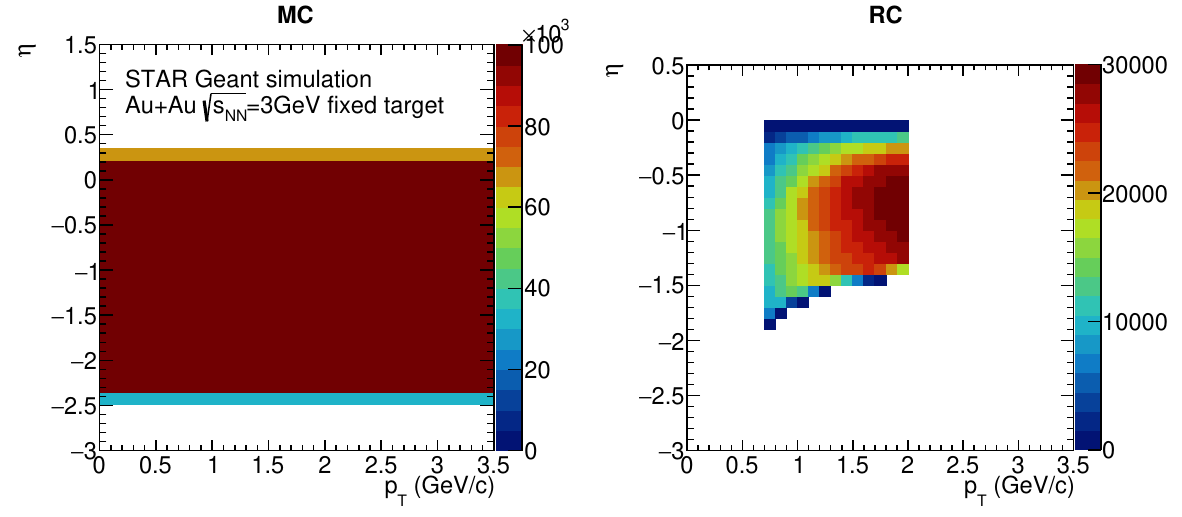}
	\caption{STAR GEANT embedding simulation for $\Lambda$ reconstruction in Au+Au fixed target collisions at $\sqrt{s_{NN}}=3$ GeV. Left: $p_T$ and $\eta$ distribution of input Monte Carlo $\Lambda$. Right: the reconstructed $\Lambda$.}
	\label{fig:Efficiency}
\end{figure}

Figure~\ref{fig:Efficiency} shows the input (MC) and reconstructed (RC) $\Lambda$ in the simulation. It can be seen that the reconstruction efficiency is higher in the center of the detector and for high $p_T$. This efficiency effect cannot be corrected by any data-driven method, and has to be corrected through detector simulations. However, the $\Lambda$ hyperon $p_T$ and $\eta$ efficiency will not distort its decay angular distribution, so it will not create a fake polarization signal but shift the averaged $P_\Lambda$ towards high $p_T$ and mid-rapidity region, if $P_\Lambda$ has $p_T$ or rapidity dependence. 

\begin{figure}[!htbp]
	\centering
	\includegraphics[width=0.48\textwidth]{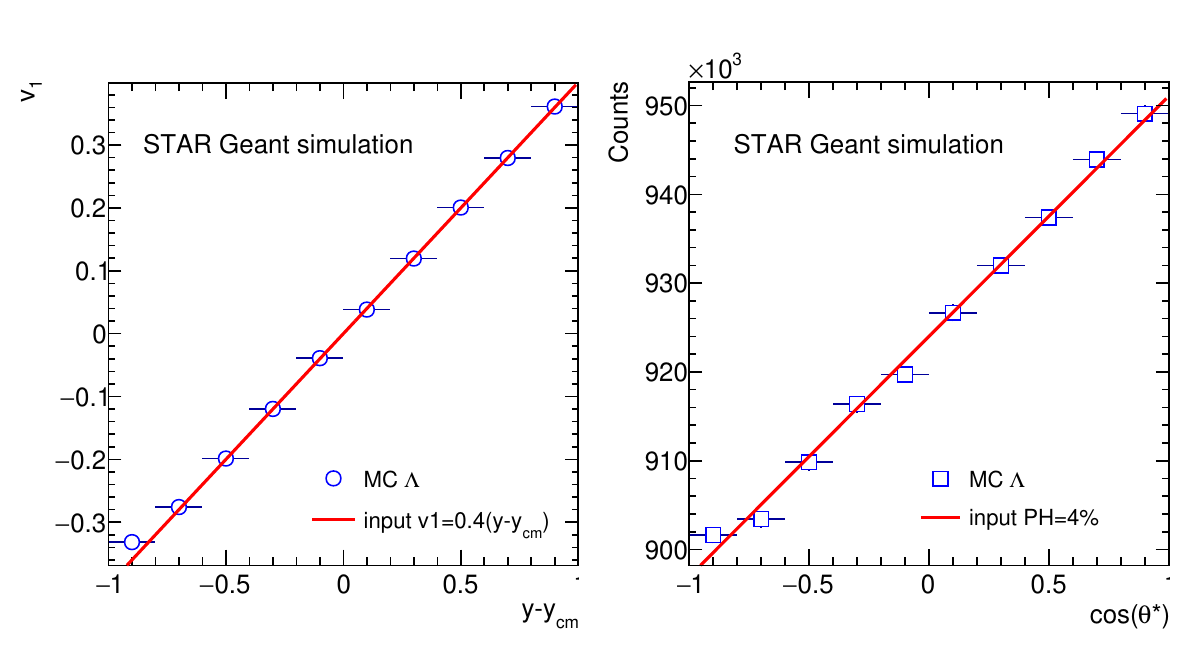}
	\caption{The directed flow as a function of rapidity in center-of-mass frame (left) and $\cos(\theta^*)$ distribution (right) of MC $\Lambda$ with input $P_\Lambda=4$\%.}
	\label{fig:input_MC}
\end{figure}

For the simulated events, the directed flow (polarization) signal is introduced by randomly throw out MC particles according to probabilities described by Eq.~\ref{eq:flow} (Eq.~\ref{eq:decay_distribution}). The directed flow is generated with a slope $dv_1/dy = 0.4$, and with no $p_T$ dependence, as illustrated in the left panel of Figure~\ref{fig:input_MC}. Right panel of Figure~\ref{fig:input_MC} shows $\cos{\theta^*}$ distribution for a polarization $P_\Lambda=4\%$. The event plane is generated randomly from 0 to $2\pi$, and is the same for the directed flow and global polarization.

\begin{figure}[!htbp]
	\centering
	\includegraphics[width=0.4\textwidth]{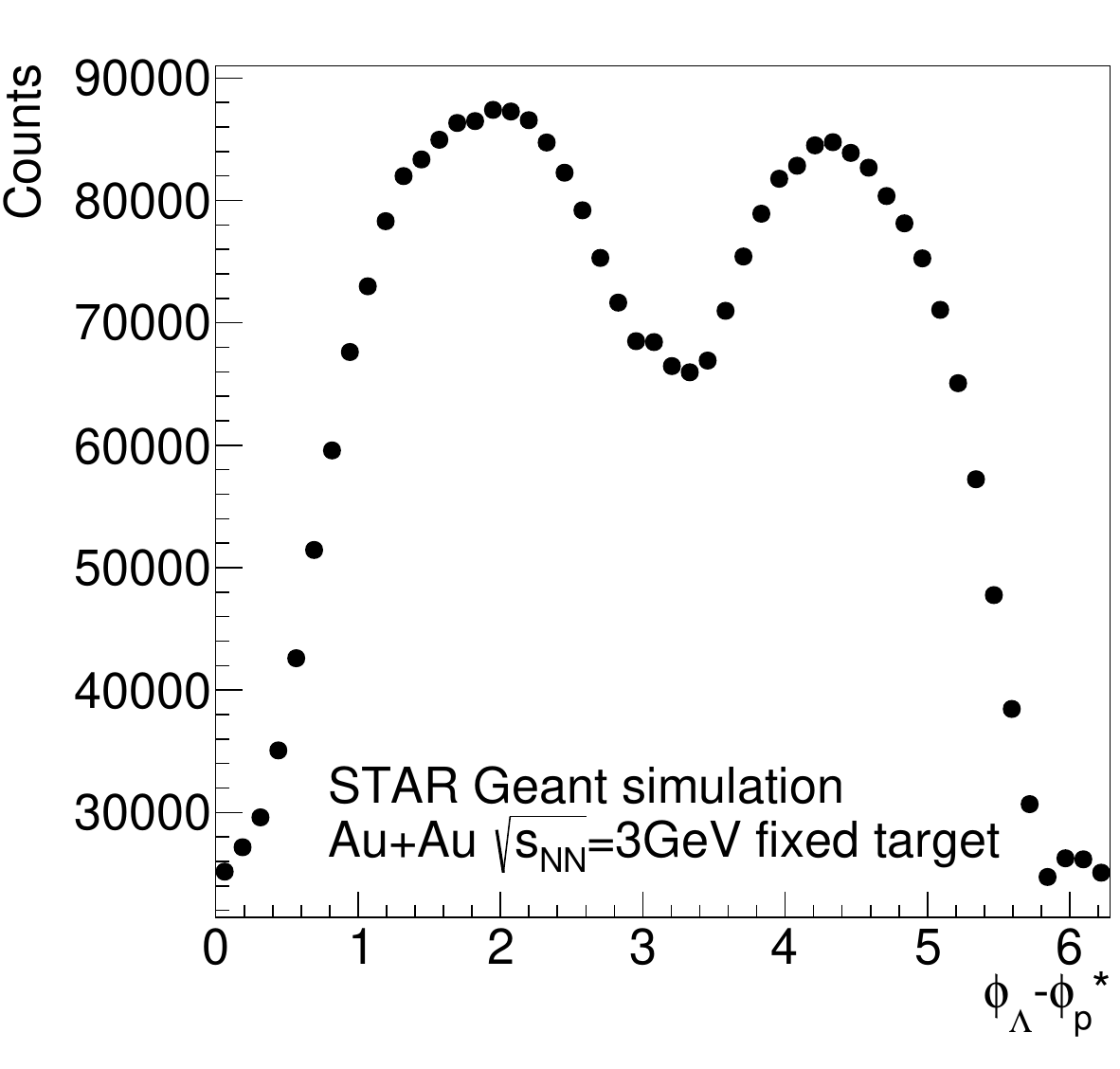}
	\caption{Counts of reconstructed $\Lambda$ hyperons as a function of $\phi_\Lambda -\phi_p^*$ in STAR GEANT embedding simulation for 20 million events.}
	\label{fig:Yields}
\end{figure}

Figure~\ref{fig:Yields} shows the $\Lambda$ counts as a function of $\phi_\Lambda - \phi_p^*$. It can be seen that the reconstruction efficiency for $\phi_\Lambda - \phi_p^*=\pi$ is higher than $0$ and $2\pi$, because the decay pions have lower $p_T^{\pi^-}$ and more difficult to be reconstructed if they are anti-parallel to $\Lambda$ moving direction. The efficiencies for proton moving perpendicular to $P_\Lambda$ are generally higher than aligned with $\Lambda$, because the later is more likely to have shared hits for the decay proton and pion (lower efficiency to construct both). The efficiency at $\pi/2$ is slightly higher than $3\pi/2$, which is because of the STAR magnet polarity. The daughter proton along $\phi_\Lambda - \phi_p^*=3\pi/2$ will cross the $\pi^-$ track when the magnetic field is along $+z$ direction, leading to a lower detection efficiency~\cite{STAR:2021beb}. 

\begin{figure}[!htbp]
	\centering
	\includegraphics[width=0.5\textwidth]{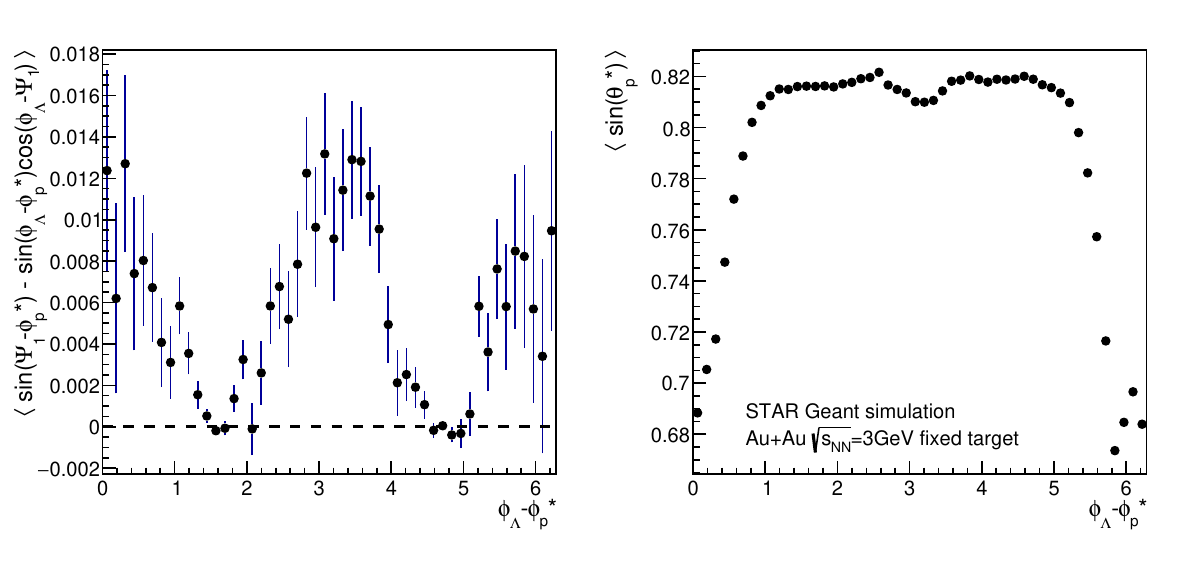}
	\caption{STAR GEANT embedding simulation for $\Lambda$ hyperons with directed flow slope $dv_1/dy=0.4$ and $P_\Lambda = 4\%$. Eq.~\ref{eq:numerator} and Eq.~\ref{eq:denominator} are shown as a function of $\phi_\Lambda -\phi_p^*$ in left and right panels, respectively.}
	\label{fig:RC_deno_numerator}
\end{figure}

The Eq.~\ref{eq:numerator} and Eq.~\ref{eq:denominator} are measured as a function of $\phi_\Lambda -\phi_p^*$, shown by Figure~\ref{fig:RC_deno_numerator}. The larger uncertainties at $0$ and $2\pi$ are because of less counts. The left panel of Figure~\ref{fig:RC_deno_numerator} shows that Eq.~\ref{eq:numerator} is 0 at $\pi/2$ and $3\pi/2$, with much smaller uncertainties. This is due to the nature of trigonometric functions, i.e. by definition it ought to be 0 according to Eq.~\ref{eq:numerator}. Both Eq.~\ref{eq:numerator} and Eq.~\ref{eq:denominator} are depends on $\phi_\Lambda -\phi_p^*$, while the shape of Eq.~\ref{eq:numerator} reflects global polarization signal and Eq.~\ref{eq:denominator} is purely from detector acceptance effect. 

\begin{figure}[!htbp]
	\centering
	\includegraphics[width=0.4\textwidth]{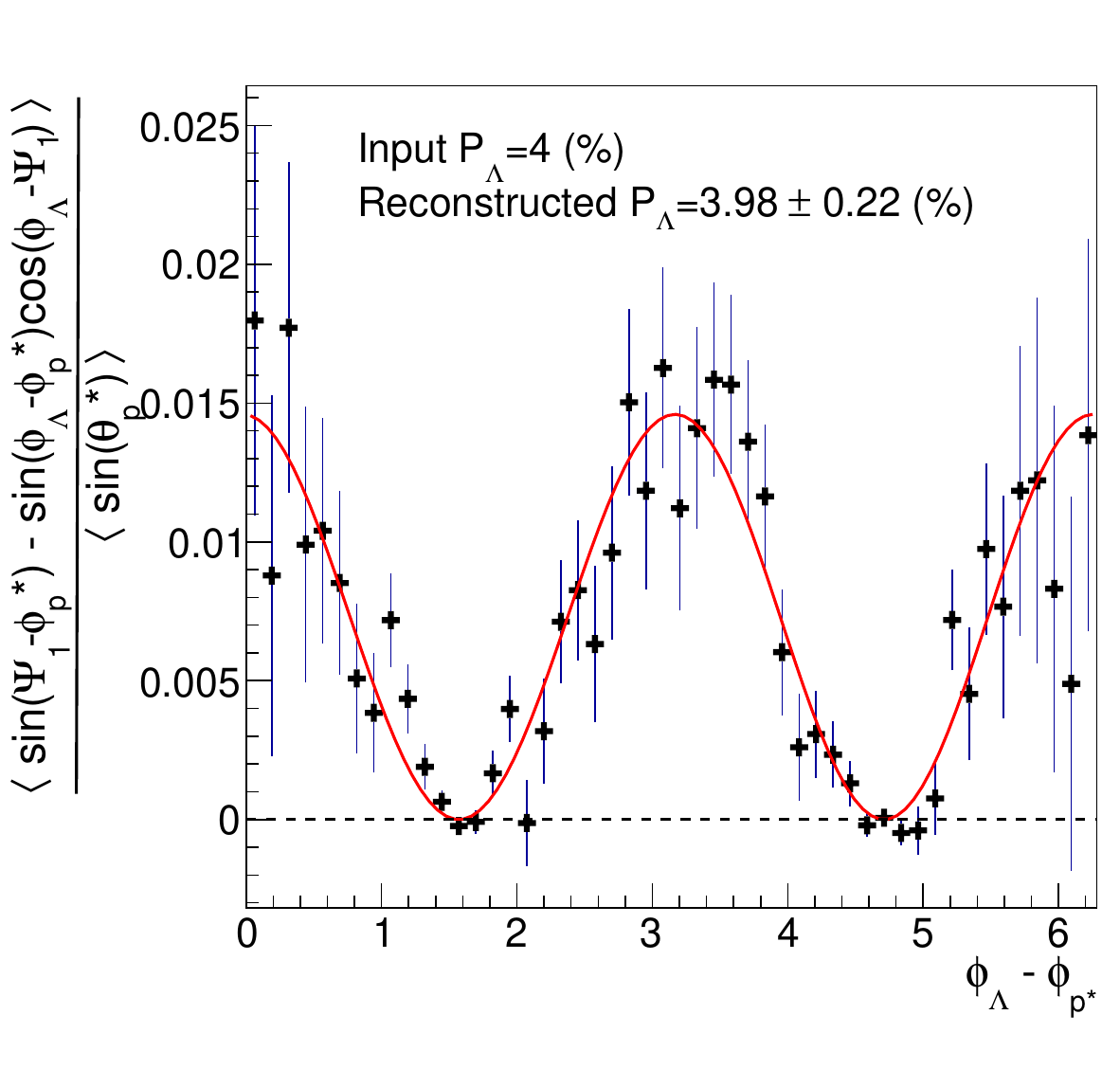}
	\caption{The ratio of Eq.~\ref{eq:numerator} over Eq.~\ref{eq:denominator} as a function of $\phi_\Lambda - \phi_p^*$. The input global polarization is 4\%, and the reconstructed polarization using fit from Eq.~\ref{eq:ratio-method} is consistent with the input.}
	\label{fig:ratio}
\end{figure}

\begin{figure*}[!htbp]
	\centering
	\includegraphics[width=0.9\textwidth]{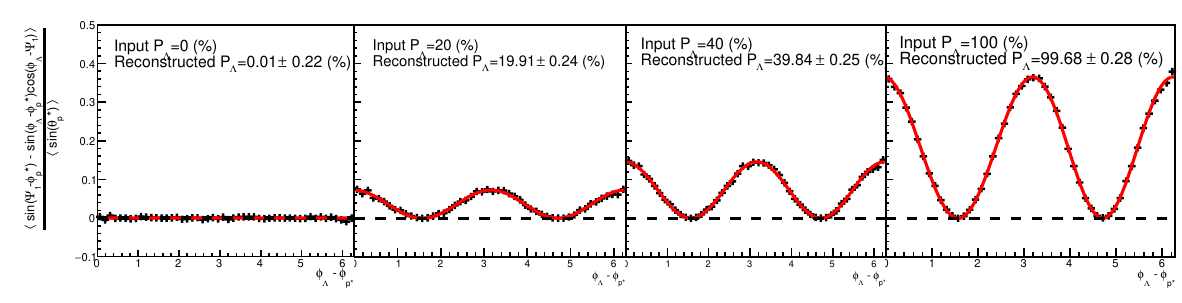}
	\caption{Illustration for the method with different global polarization strength in the simulations.}
	\label{fig:MultiPH}
\end{figure*}

Figure~\ref{fig:ratio} presents the ratio of Eq.~\ref{eq:numerator} over Eq.~\ref{eq:denominator}, and is fitted with a function described by Eq.~\ref{eq:ratio-method}. The reconstructed polarization is $3.98 \pm 0.22$ (\%) which is consistent with the input value within 1 $\sigma$ and is the typical strength of the global polarization in experiments. However, we further test the stability of the method with much stronger polarizations. Figure~\ref{fig:MultiPH} shows validation of the method for input $P_\Lambda =$0, 20\%, 40\%, and 100\%, respectively. In all these extreme cases, the input signal can be reconstructed successfully. 

\begin{figure}[!htbp]
	\centering
	\includegraphics[width=0.5\textwidth]{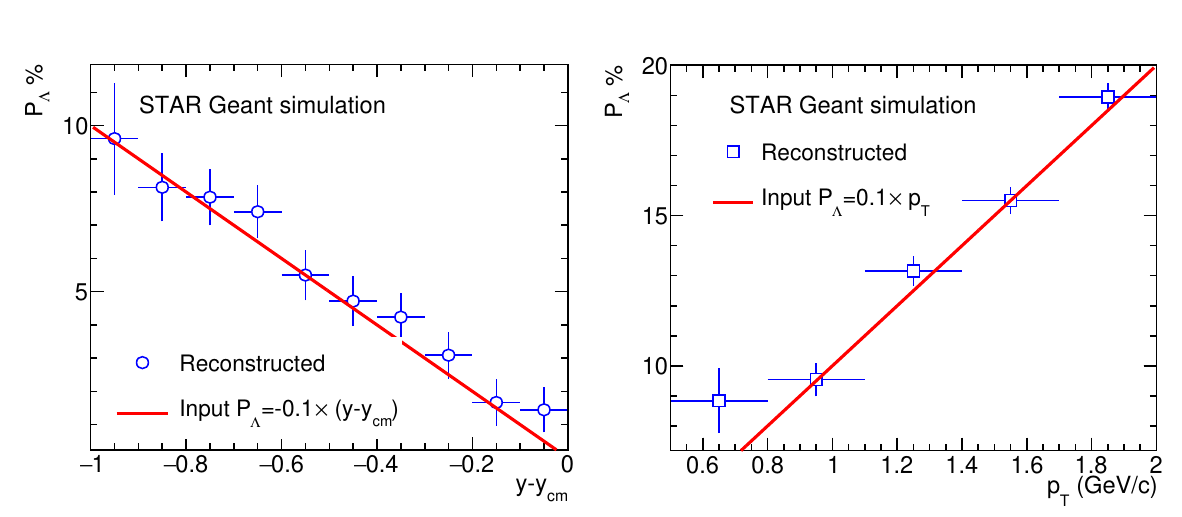}
	\caption{Global polarization measurements as functions of center-of-mass rapidity and transverse momentum. Left plot shows rapidity dependence, right plot shows $p_T$ dependence. The input values are indicated by red lines.}
	\label{fig:output_differential}
\end{figure}

In case that the polarization strongly depends on the momentum of $\rm p_\Lambda$, this method has to be applied in a narrow kinematic window. This is because the polarization $P_\Lambda$ can no longer be parameterized in Eq.~\ref{eq:numerator} if it is a function of $\Lambda$ momentum. Figure~\ref{fig:output_differential} shows when the input polarization has rapidity (left panel) or transverse momentum (right panel) dependence separately. The red lines represent input $P_\Lambda$, and the blue markers are the reconstructed polarization using Eq.~\ref{eq:ratio-method} at each rapidity or $p_T$. The polarization can be reconstructed well if it is measured differentially.

\section{Discussions}
We propose a method to measure the global polarization of $\Lambda$ hyperons under finite directed flow and with imperfect detector acceptance. However, in heavy-ion collisions, the directed flow is not the only anisotropic flow. The elliptic flow $v_2$~\cite{E895:1999ldn,STAR:2015rxv} and triangular flow $v_3$~\cite{STAR:2016vqt} can be significant at some energies, even with respect to the first-order event plane~\cite{STAR:2023duf}. After considering all the harmonics in $dN_\Lambda/d\phi_\Lambda$ distribution, Eq.~\ref{eq:numerator} is modified as
\begin{equation}
	\begin{aligned}
	&\langle\sin(\Psi_1-\phi_{p}^*) - \sin(\phi_\Lambda-\phi_p^*) \cos(\phi_\Lambda-\Psi_1)\rangle \\
	&= \frac{1}{D} \int d{\bf p_\Lambda} \int d{\bf p_p^*}  A({\bf p_\Lambda}, {\bf p_p^*)} \frac{\alpha_\Lambda P_\Lambda}{4} \sin\theta_p^* \\ 
    &\quad \times \left[ 1+(1-2v_2)\cos2(\phi_{\Lambda}-\phi_p^*) - 2v_2\sin(\phi_{\Lambda}-\phi_p^*)\right].
	\end{aligned}
	\label{eq:numerator_moreflow}
\end{equation}
It can be found that after removing directed flow contributions, there is residual elliptic flow related term. The higher harmonic flows do not have any effect on the observable. The typical $v_2$ in heavy-ion collisions above few GeV is below 10\%~\cite{STAR:2025owm}, therefore it can be safely ignored according to Eq.~\ref{eq:numerator_moreflow}.

Global polarization of $\Lambda$ hyperons is a phenomenon which describes how many $\Lambda$ are spin up or down along a common direction. So far we have only discussed how this phenomenon can be reliably measured in experiments, regardless its underling mechanisms. Recent works suggest that the transverse polarization of $\Lambda$, in couple with the directed flow, can contribute to the global polarization~\cite{Liu:2026rye}. This effect cannot be removed in our method, because such contribution is not originated from detector bias. Similarly, the polarization along $\Lambda$ helicity can also contribute to $P_y$, as along as there is a non-zero $a_1\sin(\phi_\Lambda - \Psi_1)$ term in the azimuthal angle expansion, such a term can be from the chiral vortical effect~\cite{Kharzeev:2015znc}.

\section{Summary}
In summary, we propose a method to measure the global polarization of $\Lambda$ hyperons in case of a large directed flow under asymmetric detector acceptance. This scenario usually fits the fixed target heavy-ion collisions. We provided mathematical derivations for the observables, and illustrated how the directed flow contribution is removed in the method. Using STAR GEANT embedding simulations for Au+Au fixed target collisions at $\sqrt{s_{NN}}=$ 3 GeV, we validated the method under different polarization strength. We found that with this method, the global polarization can be accurately extracted even for extreme polarization cases. We note that if the polarization strongly depends on the $\Lambda$ momentum, this method needs to be applied differentially due to mathematical requirement. We also estimated contributions from higher harmonic flows and found that only $v_2$ can have negligible effect as its magnitude is below 10\%. This study paves the way for measuring $\Lambda$ global polarization in fixed target experiments, such as STAR BES-II Fixed-Target collisions, HADES, NICA and HIRFL-CEE experiment, for a better understanding of spin dynamics in high baryon density region.

\section*{Acknowledgment}
We would like to thank the STAR collaboration and the SCDF computing facility at Brookhaven National Laboratory for providing the embedding-simulation data. We would like to thank our colleagues Nu Xu, and STAR physics working group members for many useful discussions. This work was supported by the National Key Research and Development Program of China under Contract No. 2025YFA1614300 and No. 2024YFA1610700, by the Gansu Provincial Youth Talent Program under Grant No. 2026QNGR003, by the National Natural Science Foundation of China under Grant No. 12305127, by the China Postdoctoral Science Foundation under Grant No. 2024M763212, and by the Postdoctoral Fellowship Program (Grade B) of the China Postdoctoral Science Foundation under Grant No. GZB20230737. Support from Chinese Academy of Sciences is acknowledged. 


\bibliographystyle{unsrt} 
\bibliography{example}

@article{Liang:2004ph,
    author = "Liang, Zuo-Tang and Wang, Xin-Nian",
    title = "{Globally polarized quark-gluon plasma in non-central A+A collisions}",
    eprint = "nucl-th/0410079",
    archivePrefix = "arXiv",
    reportNumber = "LBNL-56383",
    doi = "10.1103/PhysRevLett.94.102301",
    journal = "Phys. Rev. Lett.",
    volume = "94",
    pages = "102301",
    year = "2005",
    note = "[Erratum: Phys.Rev.Lett. 96, 039901 (2006)]"
}

@article{Liang:2004xn,
    author = "Liang, Zuo-Tang and Wang, Xin-Nian",
    title = "{Spin alignment of vector mesons in non-central A+A collisions}",
    eprint = "nucl-th/0411101",
    archivePrefix = "arXiv",
    reportNumber = "LBNL-56659",
    doi = "10.1016/j.physletb.2005.09.060",
    journal = "Phys. Lett. B",
    volume = "629",
    pages = "20--26",
    year = "2005"
}

@article{ParticleDataGroup:2022pth,
    author = "Workman, R. L. and others",
    collaboration = "Particle Data Group",
    title = "{Review of Particle Physics}",
    doi = "10.1093/ptep/ptac097",
    journal = "PTEP",
    volume = "2022",
    pages = "083C01",
    year = "2022"
}

@article{Voloshin:2004ha,
    author = "Voloshin, Sergei A.",
    title = "{Polarized secondary particles in unpolarized high energy hadron-hadron collisions?}",
    eprint = "nucl-th/0410089",
    archivePrefix = "arXiv",
    journal = "arXiv",
    month = "10",
    year = "2004"
}

@article{Becattini:2007sr,
    author = "Becattini, F. and Piccinini, F. and Rizzo, J.",
    title = "{Angular momentum conservation in heavy ion collisions at very high energy}",
    eprint = "0711.1253",
    archivePrefix = "arXiv",
    primaryClass = "nucl-th",
    doi = "10.1103/PhysRevC.77.024906",
    journal = "Phys. Rev. C",
    volume = "77",
    pages = "024906",
    year = "2008"
}

@article{Becattini:2013fla,
    author = "Becattini, F. and Chandra, V. and Del Zanna, L. and Grossi, E.",
    title = "{Relativistic distribution function for particles with spin at local thermodynamical equilibrium}",
    eprint = "1303.3431",
    archivePrefix = "arXiv",
    primaryClass = "nucl-th",
    doi = "10.1016/j.aop.2013.07.004",
    journal = "Annals Phys.",
    volume = "338",
    pages = "32--49",
    year = "2013"
}

@article{Becattini:2012tc,
    author = "Becattini, F.",
    title = "{Covariant statistical mechanics and the stress-energy tensor}",
    eprint = "1201.5278",
    archivePrefix = "arXiv",
    primaryClass = "gr-qc",
    doi = "10.1103/PhysRevLett.108.244502",
    journal = "Phys. Rev. Lett.",
    volume = "108",
    pages = "244502",
    year = "2012"
}

@article{Becattini:2024uha,
    author = "Becattini, Francesco and Buzzegoli, Matteo and Niida, Takafumi and Pu, Shi and Tang, Ai-Hong and Wang, Qun",
    title = "{Spin polarization in relativistic heavy-ion collisions}",
    eprint = "2402.04540",
    archivePrefix = "arXiv",
    primaryClass = "nucl-th",
    doi = "10.1142/9789811294679_0005",
    journal = "Int. J. Mod. Phys. E",
    volume = "33",
    number = "06",
    pages = "2430006",
    year = "2024"
}

@article{Karpenko:2016jyx,
    author = "Karpenko, I. and Becattini, F.",
    title = "{Study of $\Lambda $ polarization in relativistic nuclear collisions at $\sqrt{s_\mathrm {NN}}=7.7$ {\textendash}200 GeV}",
    eprint = "1610.04717",
    archivePrefix = "arXiv",
    primaryClass = "nucl-th",
    doi = "10.1140/epjc/s10052-017-4765-1",
    journal = "Eur. Phys. J. C",
    volume = "77",
    number = "4",
    pages = "213",
    year = "2017"
}

@article{Akridge:2025jgy,
    author = "Akridge, Alex and Gallimore, Daniel and Morales, Hector and Liao, Jinfeng",
    title = "{Initial baryon stopping and angular momentum in heavy-ion collisions}",
    eprint = "2504.02192",
    archivePrefix = "arXiv",
    primaryClass = "nucl-th",
    doi = "10.1103/lz5s-98rc",
    journal = "Phys. Rev. C",
    volume = "112",
    number = "6",
    pages = "064911",
    year = "2025"
}

@article{Ivanov:2025izv,
    author = "Ivanov, Yu. B.",
    title = "{Global {\ensuremath{\Lambda}} polarization in heavy-ion collisions at high baryon density}",
    eprint = "2504.12200",
    archivePrefix = "arXiv",
    primaryClass = "nucl-th",
    doi = "10.1103/qn56-fvnq",
    journal = "Phys. Rev. C",
    volume = "112",
    number = "1",
    pages = "014902",
    year = "2025"
}

@article{Ivanov:2020udj,
    author = "Ivanov, Yu B.",
    title = "{Global $\Lambda$ polarization in moderately relativistic nuclear collisions}",
    eprint = "2012.07597",
    archivePrefix = "arXiv",
    primaryClass = "nucl-th",
    doi = "10.1103/PhysRevC.103.L031903",
    journal = "Phys. Rev. C",
    volume = "103",
    number = "3",
    pages = "L031903",
    year = "2021"
}

@article{Ivanov:2019ern,
    author = "Ivanov, Yu B. and Toneev, V. D. and Soldatov, A. A.",
    title = "{Estimates of hyperon polarization in heavy-ion collisions at collision energies $\sqrt{s_{NN}}=$ 4--40 GeV}",
    eprint = "1903.05455",
    archivePrefix = "arXiv",
    primaryClass = "nucl-th",
    doi = "10.1103/PhysRevC.100.014908",
    journal = "Phys. Rev. C",
    volume = "100",
    number = "1",
    pages = "014908",
    year = "2019"
}

@article{Ivanov:2022ble,
    author = "Ivanov, Yu. B. and Soldatov, A. A.",
    title = "{Global {\ensuremath{\Lambda}} polarization in heavy-ion collisions at energies 2.4{\textendash}7.7 GeV: Effect of meson-field interaction}",
    eprint = "2201.04527",
    archivePrefix = "arXiv",
    primaryClass = "nucl-th",
    doi = "10.1103/PhysRevC.105.034915",
    journal = "Phys. Rev. C",
    volume = "105",
    number = "3",
    pages = "034915",
    year = "2022"
}

@article{Vitiuk:2019rfv,
    author = "Vitiuk, O. and Bravina, L. V. and Zabrodin, E. E.",
    title = "{Is different $\Lambda$ and $\bar \Lambda$ polarization caused by different spatio-temporal freeze-out picture?}",
    eprint = "1910.06292",
    archivePrefix = "arXiv",
    primaryClass = "hep-ph",
    doi = "10.1016/j.physletb.2020.135298",
    journal = "Phys. Lett. B",
    volume = "803",
    pages = "135298",
    year = "2020"
}

@article{Li:2017slc,
    author = "Li, Hui and Pang, Long-Gang and Wang, Qun and Xia, Xiao-Liang",
    title = "{Global $\Lambda$ polarization in heavy-ion collisions from a transport model}",
    eprint = "1704.01507",
    archivePrefix = "arXiv",
    primaryClass = "nucl-th",
    doi = "10.1103/PhysRevC.96.054908",
    journal = "Phys. Rev. C",
    volume = "96",
    number = "5",
    pages = "054908",
    year = "2017"
}

@article{Fu:2020oxj,
    author = "Fu, Baochi and Xu, Kai and Huang, Xu-Guang and Song, Huichao",
    title = "{Hydrodynamic study of hyperon spin polarization in relativistic heavy ion collisions}",
    eprint = "2011.03740",
    archivePrefix = "arXiv",
    primaryClass = "nucl-th",
    doi = "10.1103/PhysRevC.103.024903",
    journal = "Phys. Rev. C",
    volume = "103",
    number = "2",
    pages = "024903",
    year = "2021"
}

@article{Guo:2021udq,
    author = "Guo, Yu and Liao, Jinfeng and Wang, Enke and Xing, Hongxi and Zhang, Hui",
    title = "{Hyperon polarization from the vortical fluid in low-energy nuclear collisions}",
    eprint = "2105.13481",
    archivePrefix = "arXiv",
    primaryClass = "nucl-th",
    doi = "10.1103/PhysRevC.104.L041902",
    journal = "Phys. Rev. C",
    volume = "104",
    number = "4",
    pages = "L041902",
    year = "2021"
}

@article{Deng:2021miw,
    author = "Deng, Xian-Gai and Huang, Xu-Guang and Ma, Yu-Gang",
    title = "{Lambda polarization in 108Ag+108Ag and 197Au+197Au collisions around a few GeV}",
    eprint = "2109.09956",
    archivePrefix = "arXiv",
    primaryClass = "nucl-th",
    doi = "10.1016/j.physletb.2022.137560",
    journal = "Phys. Lett. B",
    volume = "835",
    pages = "137560",
    year = "2022"
}

@article{Deng:2020ygd,
    author = "Deng, Xian-Gai and Huang, Xu-Guang and Ma, Yu-Gang and Zhang, Song",
    title = "{Vorticity in low-energy heavy-ion collisions}",
    eprint = "2001.01371",
    archivePrefix = "arXiv",
    primaryClass = "nucl-th",
    doi = "10.1103/PhysRevC.101.064908",
    journal = "Phys. Rev. C",
    volume = "101",
    number = "6",
    pages = "064908",
    year = "2020"
}

@article{Becattini:2021lfq,
    author = "Becattini, Francesco and Liao, Jinfeng and Lisa, Michael",
    title = "{Strongly Interacting Matter Under Rotation: An Introduction}",
    eprint = "2102.00933",
    archivePrefix = "arXiv",
    primaryClass = "nucl-th",
    doi = "10.1007/978-3-030-71427-7_1",
    journal = "Lect. Notes Phys.",
    volume = "987",
    pages = "1--14",
    year = "2021"
}

@article{Huang:2020dtn,
    author = "Huang, Xu-Guang and Liao, Jinfeng and Wang, Qun and Xia, Xiao-Liang",
    title = "{Vorticity and Spin Polarization in Heavy Ion Collisions: Transport Models}",
    eprint = "2010.08937",
    archivePrefix = "arXiv",
    primaryClass = "nucl-th",
    doi = "10.1007/978-3-030-71427-7_9",
    journal = "Lect. Notes Phys.",
    volume = "987",
    pages = "281--308",
    year = "2021"
}

@article{Becattini:2020ngo,
    author = "Becattini, Francesco and Lisa, Michael A.",
    title = "{Polarization and Vorticity in the Quark{\textendash}Gluon Plasma}",
    eprint = "2003.03640",
    archivePrefix = "arXiv",
    primaryClass = "nucl-ex",
    doi = "10.1146/annurev-nucl-021920-095245",
    journal = "Ann. Rev. Nucl. Part. Sci.",
    volume = "70",
    pages = "395--423",
    year = "2020"
}

@article{STAR:2007ccu,
    author = "Abelev, B. I. and others",
    collaboration = "STAR",
    title = "{Global polarization measurement in Au+Au collisions}",
    eprint = "0705.1691",
    archivePrefix = "arXiv",
    primaryClass = "nucl-ex",
    reportNumber = "STAR-05-11-2007",
    doi = "10.1103/PhysRevC.76.024915",
    journal = "Phys. Rev. C",
    volume = "76",
    pages = "024915",
    year = "2007",
    note = "[Erratum: Phys.Rev.C 95, 039906 (2017)]"
}

@article{STAR:2017ckg,
    author = "Adamczyk, L. and others",
    collaboration = "STAR",
    title = "{Global $\Lambda$ hyperon polarization in nuclear collisions: evidence for the most vortical fluid}",
    eprint = "1701.06657",
    archivePrefix = "arXiv",
    primaryClass = "nucl-ex",
    doi = "10.1038/nature23004",
    journal = "Nature",
    volume = "548",
    pages = "62--65",
    year = "2017"
}

@article{STAR:2018gyt,
    author = "Adam, Jaroslav and others",
    collaboration = "STAR",
    title = "{Global polarization of $\Lambda$ hyperons in Au+Au collisions at $\sqrt{s_{_{NN}}}$ = 200 GeV}",
    eprint = "1805.04400",
    archivePrefix = "arXiv",
    primaryClass = "nucl-ex",
    doi = "10.1103/PhysRevC.98.014910",
    journal = "Phys. Rev. C",
    volume = "98",
    pages = "014910",
    year = "2018"
}

@article{STAR:2020xbm,
    author = "Adam, J. and others",
    collaboration = "STAR",
    title = "{Global Polarization of $\Xi$ and $\Omega$ Hyperons in Au+Au Collisions at $\sqrt {s_{NN}}$ = 200  GeV}",
    eprint = "2012.13601",
    archivePrefix = "arXiv",
    primaryClass = "nucl-ex",
    doi = "10.1103/PhysRevLett.126.162301",
    journal = "Phys. Rev. Lett.",
    volume = "126",
    number = "16",
    pages = "162301",
    year = "2021",
    note = "[Erratum: Phys.Rev.Lett. 131, 089901 (2023)]"
}

@article{STAR:2021beb,
    author = "Abdallah, M. S. and others",
    collaboration = "STAR",
    title = "{Global $\Lambda$-hyperon polarization in Au+Au collisions at $\sqrt {s_{NN}}$=3~GeV}",
    eprint = "2108.00044",
    archivePrefix = "arXiv",
    primaryClass = "nucl-ex",
    doi = "10.1103/PhysRevC.104.L061901",
    journal = "Phys. Rev. C",
    volume = "104",
    number = "6",
    pages = "L061901",
    year = "2021"
}

@article{STAR:2023nvo,
    author = "Abdulhamid, M. I. and others",
    collaboration = "STAR",
    title = "{Global polarization of {\ensuremath{\Lambda}} and {\ensuremath{\Lambda}}{\textasciimacron} hyperons in Au+Au collisions at sNN=19.6 and 27 GeV}",
    eprint = "2305.08705",
    archivePrefix = "arXiv",
    primaryClass = "nucl-ex",
    doi = "10.1103/PhysRevC.108.014910",
    journal = "Phys. Rev. C",
    volume = "108",
    number = "1",
    pages = "014910",
    year = "2023"
}

@article{HADES:2022enx,
    author = "Abou Yassine, R. and others",
    collaboration = "HADES",
    title = "{Measurement of global polarization of {\ensuremath{\Lambda}} hyperons in few-GeV heavy-ion collisions}",
    eprint = "2207.05160",
    archivePrefix = "arXiv",
    primaryClass = "nucl-ex",
    doi = "10.1016/j.physletb.2022.137506",
    journal = "Phys. Lett. B",
    volume = "835",
    pages = "137506",
    year = "2022"
}

@article{STAR:2025dgs,
    author = "Aboona, B. E. and others",
    collaboration = "STAR",
    title = "{Hyperon global polarization in isobar Ru+Ru and Zr+Zr collisions at sNN=200GeV}",
    eprint = "2505.05046",
    archivePrefix = "arXiv",
    primaryClass = "nucl-ex",
    doi = "10.1016/j.physletb.2025.139891",
    journal = "Phys. Lett. B",
    volume = "870",
    pages = "139891",
    year = "2025"
}

@article{STAR:2023qyt,
    author = "Abdulhamid, M. I. and others",
    collaboration = "STAR",
    title = "{Event-by-event correlations between {\ensuremath{\Lambda}}~({\ensuremath{\Lambda}}{\textasciimacron}) hyperon global polarization and handedness with charged hadron azimuthal separation in Au+Au collisions at sNN=27 GeV from STAR}",
    eprint = "2304.10037",
    archivePrefix = "arXiv",
    primaryClass = "nucl-ex",
    doi = "10.1103/PhysRevC.108.014909",
    journal = "Phys. Rev. C",
    volume = "108",
    number = "1",
    pages = "014909",
    year = "2023"
}

@article{ALICE:2019onw,
    author = "Acharya, Shreyasi and others",
    collaboration = "ALICE",
    title = "{Global polarization of $\Lambda \bar \Lambda$ hyperons in Pb-Pb collisions at $\sqrt {s_{NN}}$ = 2.76 and 5.02 TeV}",
    eprint = "1909.01281",
    archivePrefix = "arXiv",
    primaryClass = "nucl-ex",
    reportNumber = "CERN-EP-2019-173",
    doi = "10.1103/PhysRevC.101.044611",
    journal = "Phys. Rev. C",
    volume = "101",
    number = "4",
    pages = "044611",
    year = "2020",
    note = "[Erratum: Phys.Rev.C 105, 029902 (2022)]"
}

@article{STAR:2019erd,
    author = "Adam, Jaroslav and others",
    collaboration = "STAR",
    title = "{Polarization of $\Lambda$ ($\bar{\Lambda}$) hyperons along the beam direction in Au+Au collisions at $\sqrt{s_{_{NN}}}$ = 200 GeV}",
    eprint = "1905.11917",
    archivePrefix = "arXiv",
    primaryClass = "nucl-ex",
    doi = "10.1103/PhysRevLett.123.132301",
    journal = "Phys. Rev. Lett.",
    volume = "123",
    number = "13",
    pages = "132301",
    year = "2019"
}

@article{STAR:2023eck,
    author = "Abdulhamid, Muhammad and others",
    collaboration = "STAR",
    title = "{Hyperon Polarization along the Beam Direction Relative to the Second and Third Harmonic Event Planes in Isobar Collisions at sNN=200{\,}{\,}GeV}",
    eprint = "2303.09074",
    archivePrefix = "arXiv",
    primaryClass = "nucl-ex",
    doi = "10.1103/PhysRevLett.131.202301",
    journal = "Phys. Rev. Lett.",
    volume = "131",
    number = "20",
    pages = "202301",
    year = "2023"
}

@article{ALICE:2021pzu,
    author = "Acharya, Shreyasi and others",
    collaboration = "ALICE",
    title = "{Polarization of $\Lambda$ and $\bar \Lambda$ Hyperons along the Beam Direction in Pb-Pb Collisions at $\sqrt {s_{NN}}$=5.02{\,}{\,}TeV}",
    eprint = "2107.11183",
    archivePrefix = "arXiv",
    primaryClass = "nucl-ex",
    reportNumber = "CERN-EP-2021-148",
    doi = "10.1103/PhysRevLett.128.172005",
    journal = "Phys. Rev. Lett.",
    volume = "128",
    number = "17",
    pages = "172005",
    year = "2022"
}

@article{Selyuzhenkov:2006tj,
    author = "Selyuzhenkov, Ilya",
    editor = "Liss, T. M.",
    collaboration = "STAR",
    title = "{Acceptance effects in the hyperons global polarization measurement}",
    eprint = "nucl-ex/0608034",
    archivePrefix = "arXiv",
    doi = "10.1063/1.2402736",
    journal = "AIP Conf. Proc.",
    volume = "870",
    number = "1",
    pages = "712--715",
    year = "2006"
}

@article{Niida:2024ntm,
    author = "Niida, Takafumi and Voloshin, Sergei A.",
    title = "{Polarization phenomenon in heavy-ion collisions}",
    eprint = "2404.11042",
    archivePrefix = "arXiv",
    primaryClass = "nucl-ex",
    doi = "10.1142/S0218301324300108",
    journal = "Int. J. Mod. Phys. E",
    volume = "33",
    number = "09",
    pages = "2430010",
    year = "2024"
}

@article{Poskanzer:1998yz,
    author = "Poskanzer, Arthur M. and Voloshin, S. A.",
    title = "{Methods for analyzing anisotropic flow in relativistic nuclear collisions}",
    eprint = "nucl-ex/9805001",
    archivePrefix = "arXiv",
    doi = "10.1103/PhysRevC.58.1671",
    journal = "Phys. Rev. C",
    volume = "58",
    pages = "1671--1678",
    year = "1998"
}

@article{Voloshin:2008dg,
    author = "Voloshin, Sergei A. and Poskanzer, Arthur M. and Snellings, Raimond",
    editor = "Stock, R.",
    title = "{Collective phenomena in non-central nuclear collisions}",
    eprint = "0809.2949",
    archivePrefix = "arXiv",
    primaryClass = "nucl-ex",
    doi = "10.1007/978-3-642-01539-7_10",
    journal = "Landolt-Bornstein",
    volume = "23",
    pages = "293--333",
    year = "2010"
}

@article{Voronyuk:2023vyu,
    author = "Voronyuk, V. and Tsegelnik, N. S. and Kolomeitsev, E. E.",
    title = "{Hyperon global polarization in heavy-ion collisions at energies available at the JINR Nuclotron-based Ion Collider fAcility: Feed-down effects and the role of {\ensuremath{\Sigma}}0 hyperons}",
    eprint = "2305.10792",
    archivePrefix = "arXiv",
    primaryClass = "nucl-th",
    doi = "10.1103/PhysRevC.111.034907",
    journal = "Phys. Rev. C",
    volume = "111",
    number = "3",
    pages = "034907",
    year = "2025"
}

@book{Friman:2011zz,
    editor = "Friman, Bengt and Hohne, Claudia and Knoll, Jorn and Leupold, Stefan and Randrup, Jorgen and Rapp, Ralf and Senger, Peter",
    title = "{The CBM physics book: Compressed baryonic matter in laboratory experiments}",
    doi = "10.1007/978-3-642-13293-3",
    volume = "814",
    year = "2011"
}

@article{Almaalol:2022xwv,
    author = "Almaalol, D. and others",
    title = "{QCD Phase Structure and Interactions at High Baryon Density: Continuation of BES Physics Program with CBM at FAIR}",
    eprint = "2209.05009",
    archivePrefix = "arXiv",
    primaryClass = "nucl-ex",
    month = "9",
    year = "2022"
}

@article{Lu:2016htm,
    author = {L{\"u}, LiMing and Yi, Han and Xiao, ZhiGang and Shao, Ming and Zhang, Song and Xiao, GuoQing and Xu, Nu},
    title = "{Conceptual design of the HIRFL-CSR external-target experiment}",
    doi = "10.1007/s11433-016-0342-x",
    journal = "Sci. China Phys. Mech. Astron.",
    volume = "60",
    number = "1",
    pages = "012021",
    year = "2017"
}

@article{Hu:2023niz,
    author = "Hu, D. and Wang, X. and Shao, M. and Zhou, Y. and Ye, S. and Zhao, L. and Sun, Y. and Lu, J. and Xu, H.",
    title = "{Design and performance testing of a T0 detector for the CSR External-target Experiment}",
    eprint = "2304.02944",
    archivePrefix = "arXiv",
    primaryClass = "physics.ins-det",
    doi = "10.1016/j.nima.2023.168773",
    journal = "Nucl. Instrum. Meth. A",
    volume = "1057",
    pages = "168773",
    year = "2023"
}

@article{Xia:2002xpu,
    author = "Xia, J. W. and others",
    title = "{The heavy ion cooler-storage-ring project (HIRFL-CSR) at Lanzhou}",
    doi = "10.1016/S0168-9002(02)00475-8",
    journal = "Nucl. Instrum. Meth. A",
    volume = "488",
    number = "1-2",
    pages = "11--25",
    year = "2002"
}

@article{Mao:2020rlb,
    author = "Mao, L. J. and others",
    title = "{Introduction of the Heavy Ion Research Facility in Lanzhou (HIRFL)}",
    doi = "10.1088/1748-0221/15/12/T12015",
    journal = "JINST",
    volume = "15",
    number = "12",
    pages = "T12015",
    year = "2020"
}

@article{Zhou:2022pxl,
    author = "Zhou, Xiaohong and Yang, Jiancheng",
    collaboration = "HIAF project Team",
    title = "{Status of the high-intensity heavy-ion accelerator facility in China}",
    doi = "10.1007/s43673-022-00064-1",
    journal = "AAPPS Bull.",
    volume = "32",
    number = "1",
    pages = "35",
    year = "2022"
}

@article{Yang:2025sni,
    author = "Yang, Jiancheng",
    title = "{Status of the HIAF accelerator facility in China}",
    doi = "10.18429/JACoW-HIAT2025-MOY01",
    journal = "JACoW",
    volume = "HIAT2025",
    pages = "MOY01",
    year = "2025"
}

@article{Brun:1994aa,
    author = "Brun, Ren{\'e} and Bruyant, F. and Carminati, Federico and Giani, Simone and Maire, M. and McPherson, A. and Patrick, G. and Urban, L.",
    title = "{GEANT Detector Description and Simulation Tool}",
    reportNumber = "CERN-W5013, CERN-W-5013, W5013, W-5013",
    doi = "10.17181/CERN.MUHF.DMJ1",
    month = "10",
    year = "1994"
}

@techreport{Fine:2000,
  author       = {V. Fine and P. Nevski},
  title        = {OO model of the STAR detector for simulation, visualization and reconstruction},
  institution  = {Brookhaven National Laboratory},
  reportNumber = {BNL-68202},
  year         = {2000},
  url          = {https://www.osti.gov/biblio/780777}
}

@article{STAR:2001rbj,
    author = "Adler, C. and others",
    collaboration = "STAR",
    title = "{Midrapidity anti-proton to proton ratio from Au + Au collisions at s(NN)**(1/2) = 130-GeV}",
    eprint = "nucl-ex/0104022",
    archivePrefix = "arXiv",
    doi = "10.1103/PhysRevLett.86.4778",
    journal = "Phys. Rev. Lett.",
    volume = "86",
    pages = "4778",
    year = "2001",
    note = "[Erratum: Phys.Rev.Lett. 90, 119903 (2003)]"
}

@article{STAR:1997sav,
    author = "Wieman, H. and others",
    editor = "Barsotti, E. and Hoffman, E. J. and Kirsten, F. A.",
    collaboration = "STAR",
    title = "{STAR TPC at RHIC}",
    doi = "10.1109/23.603731",
    journal = "IEEE Trans. Nucl. Sci.",
    volume = "44",
    pages = "671--678",
    year = "1997"
}

@article{STAR:1999sib,
    author = "Ackermann, K. H. and others",
    editor = "Riccati, L. and Masera, M. and Vercellin, E.",
    collaboration = "STAR",
    title = "{The STAR time projection chamber}",
    doi = "10.1016/S0375-9474(99)85117-3",
    journal = "Nucl. Phys. A",
    volume = "661",
    pages = "681--685",
    year = "1999"
}

@article{Anderson:2003ur,
    author = "Anderson, M. and others",
    title = "{The Star time projection chamber: A Unique tool for studying high multiplicity events at RHIC}",
    eprint = "nucl-ex/0301015",
    archivePrefix = "arXiv",
    doi = "10.1016/S0168-9002(02)01964-2",
    journal = "Nucl. Instrum. Meth. A",
    volume = "499",
    pages = "659--678",
    year = "2003"
}

@article{STAR:2025owm,
    author = "Aboona, B. E. and others",
    collaboration = "STAR",
    title = "{Onset of Constituent Quark Number Scaling in Heavy-Ion Collisions at RHIC}",
    eprint = "2504.02531",
    archivePrefix = "arXiv",
    primaryClass = "nucl-ex",
    doi = "10.1103/2qhx-cp79",
    journal = "Phys. Rev. Lett.",
    volume = "135",
    number = "7",
    pages = "072301",
    year = "2025"
}

@article{STAR:2023duf,
    author = "Abdulhamid, Muhammad and others",
    collaboration = "STAR",
    title = "{Reaction plane correlated triangular flow in Au+Au collisions at sNN=3 GeV}",
    eprint = "2309.12610",
    archivePrefix = "arXiv",
    primaryClass = "nucl-ex",
    doi = "10.1103/PhysRevC.109.044914",
    journal = "Phys. Rev. C",
    volume = "109",
    number = "4",
    pages = "044914",
    year = "2024"
}

@article{E895:1999ldn,
    author = "Pinkenburg, C. and others",
    collaboration = "E895",
    title = "{Elliptic flow: Transition from out-of-plane to in-plane emission in Au + Au collisions}",
    eprint = "nucl-ex/9903010",
    archivePrefix = "arXiv",
    doi = "10.1103/PhysRevLett.83.1295",
    journal = "Phys. Rev. Lett.",
    volume = "83",
    pages = "1295--1298",
    year = "1999"
}

@article{STAR:2016vqt,
    author = "Adamczyk, L. and others",
    collaboration = "STAR",
    title = "{Beam Energy Dependence of the Third Harmonic of Azimuthal Correlations in Au+Au Collisions at RHIC}",
    eprint = "1601.01999",
    archivePrefix = "arXiv",
    primaryClass = "nucl-ex",
    doi = "10.1103/PhysRevLett.116.112302",
    journal = "Phys. Rev. Lett.",
    volume = "116",
    number = "11",
    pages = "112302",
    year = "2016"
}

@article{STAR:2015rxv,
    author = "Adamczyk, L. and others",
    collaboration = "STAR",
    title = "{Centrality dependence of identified particle elliptic flow in relativistic heavy ion collisions at $\sqrt{s_{NN}}$=7.7{\textendash}62.4 GeV}",
    eprint = "1509.08397",
    archivePrefix = "arXiv",
    primaryClass = "nucl-ex",
    doi = "10.1103/PhysRevC.93.014907",
    journal = "Phys. Rev. C",
    volume = "93",
    number = "1",
    pages = "014907",
    year = "2016"
}

@article{Kharzeev:2015znc,
    author = "Kharzeev, D. E. and Liao, J. and Voloshin, S. A. and Wang, G.",
    title = "{Chiral magnetic and vortical effects in high-energy nuclear collisions{\textemdash}A status report}",
    eprint = "1511.04050",
    archivePrefix = "arXiv",
    primaryClass = "hep-ph",
    doi = "10.1016/j.ppnp.2016.01.001",
    journal = "Prog. Part. Nucl. Phys.",
    volume = "88",
    pages = "1--28",
    year = "2016"
}

@article{Liu:2026rye,
    author = "Liu, Feng and Tu, Zhoudunming",
    title = "{Global $Λ$ hyperon polarization in low-energy heavy ion collisions -- a scenario without vorticity}",
    eprint = "2603.19581",
    archivePrefix = "arXiv",
    primaryClass = "hep-ph",
    month = "3",
    year = "2026"
}

@phdthesis{Gorbunov2013,
  author      = {S Gorbunov},
  title       = {On-line reconstruction algorithms for the CBM and ALICE experiments},
  type        = {doctoralthesis},
  pages       = {104, 9},
  school      = {Universit{\"a}tsbibliothek Johann Christian Senckenberg},
  year        = {2013},
}

@phdthesis{Zyzak2016,
  author      = {M Zyzak},
  title       = {Online selection of short-lived particles on many-core computer architectures in the CBM experiment at FAIR},
  type        = {doctoralthesis},
  pages       = {104, 9},
  school      = {Johann Wolfgang Goethe Universit{\"a}t},
  year        = {2016},
}

@article{Kisel:2018nvd,
    author = "Kisel, Ivan",
    editor = "Aichelin, Joerg and Bellwied, Rene and Bratkovskaya, Elena and Timmins, Anthony",
    collaboration = "CBM",
    title = "{Event Topology Reconstruction in the CBM Experiment}",
    doi = "10.1088/1742-6596/1070/1/012015",
    journal = "J. Phys. Conf. Ser.",
    volume = "1070",
    number = "1",
    pages = "012015",
    year = "2018"
}

@article{Ju:2023xvg,
    author = "Ju, Xin-Yue and others",
    title = "{Applying the Kalman filter particle method to strange and open charm hadron reconstruction in the STAR experiment}",
    doi = "10.1007/s41365-023-01320-1",
    journal = "Nucl. Sci. Tech.",
    volume = "34",
    number = "10",
    pages = "158",
    year = "2023"
}

@article{Chen:2024aom,
    author = "Chen, Jinhui and others",
    title = "{Properties of the QCD matter: review of selected results from the relativistic heavy ion collider beam energy scan (RHIC BES) program}",
    eprint = "2407.02935",
    archivePrefix = "arXiv",
    primaryClass = "nucl-ex",
    doi = "10.1007/s41365-024-01591-2",
    journal = "Nucl. Sci. Tech.",
    volume = "35",
    number = "12",
    pages = "214",
    year = "2024"
}






\end{document}